\documentclass[12pt]{article}

        \usepackage{amsfonts}
        \usepackage{amssymb}

\def\be{\begin{eqnarray}}
\def\ee{\end{eqnarray}}
\def\bee{\begin{eqnarray*}}
\def\eee{\end{eqnarray*}}

\newtheorem{thm}{Theorem}

\newtheorem{lemma}[thm]{Lemma}

        \def\tr{\hbox{Tr}}
        
      \def\Hil{{\cal H}}

\def\Tr{{\rm Tr}}

          \title{
Additivity for unital qubit channels.}

        \author{Christopher King \\ Department of
        Mathematics \\ Northeastern University \\
 Boston, MA 02115  \\ {\normalsize king@neu.edu}
}
\begin{document}

\maketitle

\begin{abstract}
Additivity of the Holevo capacity is proved for product channels,
under the condition that one of the channels is
a unital qubit channel, with the other completely arbitrary.
As a byproduct this proves that the Holevo bound is the ultimate
information capacity of such qubit channels (assuming no
prior entanglement between sender and receiver).
Additivity of minimal entropy and multiplicativity of
$p$-norms are also proved under the same assumptions.
The proof relies on a new bound for the $p$-norm of an output
state from the phase-damping channel.
\end{abstract}

\pagebreak



\section{Introduction and statement of results}

There are several outstanding conjectures concerning product channels.
These all involve the question of finding the ``best'' state or
set of states to 
transmit through a product channel, using some measure of performance
at the output to determine ``best''. 
In this paper we will consider three well-known
measures of performance: the maximal non-commutative $p$-norm of 
an output state (this was introduced by Amosov, Holevo and Werner \cite{AHW},
who called it the `maximal output purity' of the channel);
the minimal entropy of an output state; and the Holevo quantity
${\chi}^{*}$, which is a measure of the channel's capacity for
transmitting classical information. An obvious candidate for the ``best''
set of states in each case is constructed by taking the product of the corresponding best states
for the individual channels. For the three performance measures described above,
the outstanding conjecture is that this procedure indeed produces the optimal state.
That is, the ``best'' states to send through the channel are always product states --
there is no advantage in using entangled states.
This predicts that the maximal $p$-norm is multiplicative
for product channels, and that the minimal entropy and Holevo quantity
are additive.
At this time, it seems fair to say that there is no good understanding
of why these conjectures should be true (or even whether they are true in every case).

\medskip
The conjectures have been verified numerically for products of low dimensional
channels. However numerical testing becomes difficult when the dimensions of the
state spaces are large. In part, this is because the allowed input states for the product
channel include all entangled states in addition to the product states.
So it seems to be necessary to develop an analytical method to investigate 
the conjectures, and that is the overall goal of the work reported in this paper.

\medskip
Recall that a channel $\Phi$ on a Hilbert space $\Hil$ 
is a completely positive, trace-preserving map
on the algebra of observables on $\Hil$.
When $\Hil = {\bf C}^2$ we will call $\Phi$ a {\it qubit channel}.
Qubit channels play an important role in quantum information theory,
because many applications involve the manipulation and 
entanglement of qubit states.
If the channel satisfies $\Phi(I) = I$, so that it maps the identity to
itself, then $\Phi$ is a {\it unital} channel. Examples of unital qubit 
channels are the depolarizing channel, the phase-damping channel,
and the two-Pauli channel of Bennett, Fuchs and Smolin \cite{BFS}.
The unital qubit channels provide a very useful laboratory for
testing analytical approaches to the conjectures. This is because they
are parametrized by three real numbers (up to unitary equivalence),
and the geometry of this set of parameters is well understood.
In this paper we will use detailed properties of
this class of channels to derive some new bounds which lead to proofs
of the conjectures in some special cases. Hopefully these results will 
provide clues about how to proceed in the general case.

\medskip
The results in this paper concern product channels $\Omega \otimes \Phi$
where $\Phi$ is a {\it unital qubit channel}, and $\Omega$ is {\it completely
arbitrary}. For such channels we are able to establish the conjectures
described above, namely that the three performance measures are optimized on
product states of the channel.
The main ingredient in the proof is a new inequality for the $p$-norm of an output
state from the half-noisy channel $I \otimes \Phi$.
The proof of this bound uses details of the classification of
unital qubit channels \cite{KR}, and does not obviously extend to
other classes of channels. In essence, it uses convexity and symmetry arguments
to reduce the bound to the case of the {\it phase-damping channel}.
The phase-damping channel (defined below in (\ref{ph-damp})) is a one-parameter
family of unital channels ${\Psi}_{\lambda}$ which has been used
as a model for decoherence in a two-state system. 
The proof of the bound for the
channel $I \otimes {\Psi}_{\lambda}$ is based on a result of Epstein \cite{Ep}
concerning concavity of a certain trace function.
The bound for the half-noisy channel $I \otimes \Phi$
is then enough to prove our results for the
product channel $\Omega  \otimes \Phi$.

\medskip
Before stating precisely our results we review the three performance measures for a channel
that are used here.
First, for any $p \geq 1$ the maximal $p$-norm of the channel $\Phi$ is defined by
\be\label{def:l_p}
{\nu}_{p}(\Phi) = \sup_{\rho} \, || \Phi(\rho) ||_{p},
\ee
where the $\sup$ runs over states and where the $p$-norm of a positive matrix
$A$ is defined by
\be
|| A ||_{p} = \big( \Tr A^{p} \big)^{1 \over p}
\ee

\medskip

\noindent Second, the minimal entropy of the channel $\Phi$ is defined by
\be\label{def:S_min}
S_{\rm min}(\Phi) = \inf_{\rho} S(\Phi(\rho))
\ee
where $S(\rho) = - \Tr \rho \log \rho$ is the von Neumann entropy of the state $\rho$.
\medskip

\noindent Third, the Holevo capacity of $\Phi$ is defined by
\be\label{HSW}
{\chi}^{*}(\Phi) = \sup_{\pi, \, \rho} \bigg[S \big(\sum {\pi}_i \Phi({\rho}_{i})\big)
- \sum {\pi}_i S(\Phi({\rho}_{i})) \bigg],
\ee
where the $\sup$ runs over all probability distributions $\{{\pi}_{i}\}$
and collections of states $\{{\rho}_{i}\}$ on $\Hil$.

\medskip
\begin{thm}\label{thm1}
Let $\Phi$ be a unital qubit channel. Then for any channel $\Omega$,
\be\label{eq1thm1}
\hskip1in {\nu}_{p}(\Omega \otimes \Phi) = {\nu}_{p}(\Omega) \, 
{\nu}_{p}(\Phi), \quad\quad \mbox{for any}\quad p \geq 1 
\ee
\be\label{eq2thm1}
S_{\rm min}(\Omega \otimes \Phi) = S_{\rm min}(\Omega) +
S_{\rm min}(\Phi)
\ee
\be\label{eq3thm1}
{\chi}^{*}(\Omega \otimes \Phi) = {\chi}^{*}(\Omega) +
{\chi}^{*}(\Phi)
\ee
\end{thm}
\bigskip

\bigskip
\noindent{\it Remark 1}.
Results related to Theorem \ref{thm1} have been proven before.
Several authors have proven the results for the half-noisy channel
$\Omega \otimes I$ \cite{AHW}, \cite{Fu2}, \cite{SW2}.
Holevo  proved (\ref{eq3thm1}) when both $\Omega$ and $\Phi$ are QC or CQ
channels \cite{H2}. In \cite{K},  (\ref{eq1thm1}), (\ref{eq2thm1}) and (\ref{eq3thm1})
were proven for any channel $\Omega$, when $\Phi$ is either a QC or CQ channel.
Bruss et al proved (\ref{eq3thm1}) when both $\Omega$  and $\Phi$ are 
depolarizing qubit channels \cite{BFMP}. Amosov and Holevo proved 
(\ref{eq1thm1}) for integer values of $p$ when both 
$\Omega$  and $\Phi$ are products of
depolarizing channels \cite{AH}. King and Ruskai presented strong evidence for
(\ref{eq3thm1}) when both $\Omega$  and $\Phi$ are unital qubit channels
\cite{KR}. In \cite{K} it was shown that (\ref{eq1thm1}) holds for any $\Omega$
when $p$ is integer and $\Phi$ is a unital qubit map, or when $p=2$ and
$\Phi$ is any qubit map.

\medskip
\noindent{\it Remark 2}.
The well-known Holevo-Schumacher-Westmoreland theorem \cite{H1}, \cite{SW1}
shows that ${\chi}^*(\Phi)$ is the best rate for transmission of classical
information through the channel $\Phi$ when product states are used at the
input (and possibly entangled measurements are used at the output).
As a consequence,  the ultimate capacity of a quantum channel $\Phi$ 
for faithful transmission of classical
information (without prior entanglement) is given by
\be\label{def:C_ult}
C_{\rm ult}(\Phi) = \lim_{n \rightarrow \infty}
{1 \over n} \, {\chi}^{*}({\Phi}^{\otimes n})
\ee
It follows from (\ref{eq3thm1}) that for any unital qubit
channel $\Phi$ this ultimate capacity is
\be
C_{\rm ult}(\Phi) = {\chi}^{*}(\Phi)
\ee

\medskip
\medskip

As mentioned in the introduction, our proof uses a new bound for the phase-damping
channel ${\Psi}_{\lambda}$. This one-parameter family of unital qubit channels 
is defined as follows:
\be\label{ph-damp}
{\Psi}_{\lambda}(r) = {\Psi}_{\lambda} \left(\matrix{r_{11} & r_{12} \cr
r_{21} & r_{22} }\right) = 
\left(\matrix{r_{11} & \lambda r_{12} \cr
\lambda r_{21} & r_{22} }\right)
\ee
where $-1 \leq \lambda \leq 1$. So ${\Psi}_{\lambda}$ reduces the off-diagonal entries
of $r$ and leaves unchanged the diagonal entries.
In order to state our new bound, let $\rho$ be a state on ${\bf C}^K \otimes {\bf C}^2$
for some $K$. Then $\rho$ can be written in the form
\be\label{rho}
\rho = X \otimes I + \sum_{i=1}^{3}  Y_{i} \otimes {\sigma}_{i}
= \left(\matrix{X + Y_{3} & Y_{1} - i Y_{2} \cr Y_{1} + i Y_{2} & X - Y_{3} }\right)
\ee 
where $X, Y_{i}$ are $K \times K$ matrices, with $\tr X = 1/2$.
Also the positivity of $\rho$ implies that
\be
X + Y_{3} \geq 0, \quad\quad
X - Y_{3} \geq 0
\ee
It follows from (\ref{ph-damp}) that
\be\label{Psi-rho}
(I \otimes {\Psi}_{\lambda}) (\rho) 
= \left(\matrix{X + Y_{3} & \lambda \, (Y_{1} - i Y_{2}) \cr 
\lambda \, (Y_{1} + i Y_{2}) & X - Y_{3} }\right)
\ee 
Define
\be\label{def:m_p}
m_{p}(x) = 
 \bigg[ \bigg({1 + x \over 2}\bigg)^{p} + 
\bigg({1 - x \over 2}\bigg)^{p} \bigg]^{1/p}
\ee

\medskip
\begin{thm}\label{thm2}
Let $\rho$ be a state on ${\bf C}^K \otimes {\bf C}^2$ written in the form
(\ref{rho}), and let ${\Psi}_{\lambda}$ be the phase-damping channel defined in
(\ref{ph-damp}). Then for all $p \geq 1$
\be\label{bound-Psi}
|| (I \otimes {\Psi}_{\lambda}) (\rho) ||_{p} \leq
2 \, m_{p}(\lambda) \, 
\bigg[ {1 \over 2} \tr \big( X + Y_{3} \big)^{p}
+ {1 \over 2} \tr \big( X - Y_{3} \big)^{p}
\bigg]^{1/p}
\ee
\end{thm}
 
\bigskip
\noindent Notice that setting $\lambda = 0$ in (\ref{Psi-rho}) gives
\be\label{Psi-rho-0}
(I \otimes {\Psi}_{0}) (\rho) 
= \left(\matrix{X + Y_{3} & 0 \cr 
0 & X - Y_{3} }\right)
\ee 
and that setting $x = 0$ in (\ref{def:m_p}) gives $m_{p}(0) = 2^{-1 + 1/p}$.
Therefore the bound (\ref{bound-Psi}) can be re-stated as follows:
\be\label{restate}
|| (I \otimes {\Psi}_{\lambda}) (\rho) ||_{p} \leq
m_{p}(0)^{-1} \, m_{p}(\lambda) \, 
|| (I \otimes {\Psi}_{0}) (\rho) ||_{p}
\ee

\medskip
We will use detailed knowledge of the set of unital
qubit maps, together with convexity and symmetry arguments,
to derive a bound similar to
(\ref{restate}) for the half-noisy channel $I \otimes \Phi$,
where $\Phi$ is any unital qubit channel. This bound is the
content of Theorem \ref{thm3} below. The bound will 
involve states of the form 
$(I \otimes U) \rho (I \otimes U^{*})$
where $U$ is a $2 \times 2$ unitary matrix.
Using the notation in (\ref{rho}), 
this can be written in $2 \times 2$ block form as
\be\label{rho-diag}
(I \otimes U) \rho (I \otimes U^{*})
= \left(\matrix{X + Y & Z \cr Z^{*} & X - Y}\right)
\ee
where $Y$ is some linear combination of the three matrices
$Y_{1}, Y_{2}, Y_{3}$. 
A key part of the theorem is that the unitary matrices $U$
that appear in the bound can be chosen
so that $\tr Y = 0$ in (\ref{rho-diag}). If we define
$r = {\tr}_{1} \rho$ to be the $2 \times 2$ reduced density matrix of $\rho$,
then this condition can be written as
\be
\tr \, ({\sigma}_{3} U r U^{*} ) = 0
\ee

\medskip
\begin{thm}\label{thm3}
Let $\Phi$ be a unital qubit channel. Let
$\rho$ be a state on ${\bf C}^K \otimes {\bf C}^2$, and let 
$r = {\tr}_{1} \rho$ be its reduced density matrix. 
Then there exist (i) an
integer $N \geq 1$, (ii) positive numbers $\{c_{1}, \dots, c_{N}\}$ satisfying
$\sum c_{i} = 1$, and (iii) unitary $2 \times 2$ matrices $U_{1}, \dots, U_{N}$
satisfying $\tr \,\big({\sigma}_{3} U_{i} r U_{i}^{*}\big) = 0$ for 
$i= 1,\dots,N$, such that for all $p \geq 1$
\be\label{ineq}
||(I \otimes \Phi) (\rho) \big)||_{p} \leq
m_{p}(0)^{-1} \, {\nu}_{p}(\Phi) \, \sum_{i=1}^{N} c_{i} \,
|| \big(I \otimes {\Psi}_{0}\big) \, \big((I \otimes U_{i}) 
\rho (I \otimes U_{i}^{*})\big) ||_{p}
\ee
\end{thm}

\section{Proof of Theorem \ref{thm1}}
We will assume in this section that Theorem \ref{thm3} holds,
and use it to deduce Theorem \ref{thm1}. The proof of Theorem \ref{thm3}
appears in the next section.

\medskip
First we prove the multiplicativity of the $p$-norm. It is sufficient to prove
that for any state $\tau$,
\be\label{suff1}
|| (\Omega \otimes \Phi) (\tau) ||_{p} \leq {\nu}_{p}(\Omega) \, {\nu}_{p}(\Phi)
\ee
since the inequality in the 
other direction follows trivially by restricting to product states.
Let
\be\label{rho-tau}
\rho = (\Omega \otimes I) (\tau)
\ee 
so that 
\be\label{relate}
(\Omega \otimes \Phi) (\tau) = (I \otimes \Phi) (\rho)
\ee
We use the same notation as in (\ref{rho}), and write
\be\label{def:tau,rho}
\tau = A \otimes I + \sum B_{i} \otimes {\sigma}_{i}, \quad\quad
\rho  = X \otimes I + \sum Y_{i} \otimes {\sigma}_{i}
\ee
It follows that $X = \Omega (A)$ and $Y_{i} = \Omega (B_{i})$. 
Suppose that one of the states appearing inside the sum 
on the right side of (\ref{ineq})  is
\be\label{eqn1}
(I \otimes U) \rho (I \otimes U^{*})
= \left(\matrix{X + Y & Z \cr Z^{*} & X - Y}\right)
\ee
where $Y$ is some linear combination of the matrices
$Y_{1}, Y_{2}, Y_{3}$ in (\ref{def:tau,rho}). The positivity of the left side of
(\ref{eqn1}) implies that $X \pm Y \geq 0$,
and the condition $\tr ({\sigma}_{3} U r U^{*}) = 0$ from Theorem \ref{thm3} 
means that
\be\label{tr=0}
\tr \, Y =  0
\ee
The relation (\ref{rho-tau}) implies that
\be\label{outputs}
X + Y = \Omega( A + B)
\ee
where $B$ is the same linear combination of the matrices
$B_{1}, B_{2}, B_{3}$ in (\ref{def:tau,rho}). 
Since $\Omega$ is trace-preserving, it follows from (\ref{tr=0})  that
\be\label{tr=0'}
\tr \, B = 0
\ee
Since $\tau$ in (\ref{def:tau,rho}) is a state, it follows that
$\tr A = 1/2$, and therefore (\ref{tr=0'}) implies that we can define the two states 
\be\label{def:alpha,beta}
\alpha = 2 (A + B) , \quad\quad
\beta = 2 (A - B)
\ee
Hence applying $(I \otimes {\Psi}_{0})$ to (\ref{eqn1}) and using
(\ref{relate}) gives
\be\label{eqn1'}
\big(I \otimes {\Psi}_{0}\big) \, \big(
(I \otimes U) \rho (I \otimes U^{*})\big)
= {1 \over 2} \,\left(\matrix{\Omega(\alpha) & 0 \cr 0 & \Omega(\beta)}\right)
\ee
Therefore the term on the right side of (\ref{ineq}) corresponding
to the state (\ref{eqn1}) is equal to
\be\label{eqn2}
|| \big(I \otimes {\Psi}_{0}) \, \big(
(I \otimes U) \rho (I \otimes U^{*})\big) ||_{p}
= {1 \over 2} \, \bigg[  \tr\, \Omega(\alpha)^{p} +
 \tr\, \Omega(\beta)^{p} \bigg]^{1/p}
\ee
The definition of the $p$-norm of $\Omega$ implies that
\be
\tr\, \Omega(\alpha)^{p} \leq \bigg( {\nu}_{p}(\Omega) \bigg)^{p},
\quad\quad
\tr \,\Omega(\beta)^{p} \leq \bigg( {\nu}_{p}(\Omega) \bigg)^{p}
\ee
Hence (\ref{eqn2}) yields the bound
\be
|| (I \otimes {\Psi}_{0}) ((I \otimes U) \rho (I \otimes U^{*})) ||_{p}
\leq 2^{-1 + 1/p} \,\, 
{\nu}_{p}(\Omega) =
m_{p}(0) \, \, {\nu}_{p}(\Omega)
\ee
We can repeat the same argument for all terms on the right side of (\ref{ineq}),
leading to the bound
\be\label{eqn3}
|| (I \otimes \Phi) (\rho) ||_{p} \leq
m_{p}(0)^{-1} \, {\nu}_{p}(\Phi) \, \sum_{i=1}^{N} c_{i} \,
\bigg[ m_{p}(0) \, {\nu}_{p}(\Omega) \bigg]
= {\nu}_{p}(\Phi) \, {\nu}_{p}(\Omega)
\ee
Combining (\ref{eqn3}) with (\ref{relate}) establishes the bound
(\ref{suff1}), and hence proves (\ref{eq1thm1}).

Turning now to the minimal entropy equality (\ref{eq2thm1}), this follows
immediately from (\ref{eq1thm1}) by taking the derivative at $p =1$, since 
from the easily established relation
\be\label{getS}
{d \over dp} \bigg( || \rho ||_{p}  \bigg)_{p=1} =
- S(\rho)
\ee
it follows that for any channel $\Omega$
\be\label{s_min}
{d \over dp} \bigg( {\nu}_{p}(\Omega)  \bigg)_{p=1}
= - S_{\rm min}(\Omega)
\ee

Next we turn to the additivity of the Holevo capacity (\ref{HSW}).
To establish this we use the representation of Ohya, Petz and Watanabe \cite{OPW}
and Schumacher-Westmoreland \cite{SW2}, and follow the method described
in \cite{K}. Denote the relative entropy of
states
$\rho$ and $\omega$ by
\be
S( \rho \, | \, \omega ) = \tr \rho ( \log \rho - \log \omega )
\ee
Then the OPWSW representation is
\be\label{OPWSW}
{\chi}^{*}(\Omega) = \inf_{\rho} \sup_{\omega} S \big( \Omega (\omega) \,|\,
\Omega(\rho) \big)
\ee
The state that achieves the infimum in (\ref{OPWSW}) 
is the optimal average output state from the 
channel, and we denote this by $\rho_{\Omega}$. For a unital qubit channel
$\Phi$, the optimal output state is
\be\label{opt}
\rho_{\Phi} = {1 \over 2} \, I,
\ee
and hence it follows that
\be\label{chi_Phi}
{\chi}^{*}(\Phi) = \sup_{\omega} \big( - S(\Phi(\omega)) + \log 2 \big)
= \log 2 \,\, - \,\, S_{\rm min}(\Phi)
\ee

Our goal is to show that 
\be\label{eqn4}
{\chi}^{*}(\Omega \otimes \Phi) \leq {\chi}^{*}(\Omega) + {\chi}^{*}(\Phi)
\ee
(the inequality in the other direction is trivial). From
(\ref{OPWSW}) it follows that
\be
{\chi}^{*}(\Omega \otimes \Phi) \leq
\sup_{\tau} S( (\Omega \otimes \Phi) (\tau) \,|\, {\rho}_{\Omega} \otimes 
{\rho}_{\Phi} )
\ee 
and hence to prove (\ref{eqn4}) it is sufficient to prove that for any state
$\tau$,
\be\label{eqn5}
S( (\Omega \otimes \Phi) (\tau) \,|\, {\rho}_{\Omega} \otimes 
{\rho}_{\Phi} ) \leq {\chi}^{*}(\Omega) + {\chi}^{*}(\Phi)
\ee
Denote the reduced density matrix of $\tau$ by
\be\label{def:omega}
\omega = {\tr}_{2} (\tau),
\ee 
where ${\tr}_{2}$ is the trace over the second factor.
Using (\ref{opt}) and (\ref{chi_Phi}) reduces (\ref{eqn5}) to the inequality
\be\label{suff3}
S_{\rm min}(\Phi) - S( (\Omega \otimes \Phi) (\tau) ) - 
\tr \Omega( \omega) \log ({\rho}_{\Omega})
\leq {\chi}^{*}(\Omega) 
\ee

In order to establish (\ref{suff3}) we will take the limit
$p \rightarrow 1$ in the inequality (\ref{ineq}).
Following the notation of (\ref{def:alpha,beta}) and (\ref{eqn1'}),
and recalling (\ref{rho-tau}),
the $i^{\rm th}$ term on the right
side of (\ref{ineq}) can be written as
\be
\big(I \otimes {\Psi}_{0}\big) \, \big(
(I \otimes U_{i}) \rho (I \otimes U_{i}^{*})\big)
= {1 \over 2} \,\left(\matrix{\Omega({\alpha}^{(i)}) & 0 \cr 
0 & \Omega({\beta}^{(i)})}\right)
\ee
where ${\alpha}^{(i)}$ and ${\beta}^{(i)}$ are states satisfying
\be\label{sum}
{\alpha}^{(i)} + {\beta}^{(i)} = 4 \, A = 2 \, \omega
\ee
Then the inequality (\ref{ineq}) can be written
\be
|| (\Omega \otimes \Phi) (\tau) ||_{p} \leq
{\nu}_{p}(\Phi) \, \sum_{i=1}^{N} c_{i} \,
 \bigg[ {1 \over 2} \, \tr \big( \Omega({\alpha}^{(i)}) \big)^{p}
+ {1 \over 2} \, \tr \big( \Omega({\beta}^{(i)}) \big)^{p}
\bigg]^{1/p}
\ee
This becomes an equality at $p=1$, hence taking the derivative at $p=1$ 
and using (\ref{getS}) and (\ref{s_min}) gives the bound
\be
S \big( (\Omega \otimes \Phi) (\tau) \big) \geq
S_{\rm min}(\Phi) + \sum_{i=1}^{N} c_{i} \,
\bigg[ {1 \over 2} S \big( \Omega({\alpha}^{(i)}) \big)
+ {1 \over 2} S \big( \Omega({\beta}^{(i)}) \big)
\bigg]
\ee
Comparing with the left side of (\ref{suff3}) it is sufficient to prove
\be\label{suff4}
-  \sum_{i=1}^{N} c_{i} \,
\bigg[ {1 \over 2} S \big( \Omega({\alpha}^{(i)}) \big)
+ {1 \over 2} S \big( \Omega({\beta}^{(i)}) \big)
\bigg] - \tr \Omega( \omega) \log ({\rho}_{\Omega})
\leq {\chi}^{*}(\Omega)
\ee
Using the relation (\ref{sum}) and the condition $\sum c_{i} = 1$ gives
\be
\omega = \sum_{i=1}^{N} c_{i} \,
\bigg[{1 \over 2} {\alpha}^{(i)}
+ {1 \over 2}  {\beta}^{(i)} \bigg]
\ee
Hence the left side of (\ref{suff4}) is equal to
\be\label{suff5}
\sum_{i=1}^{N} c_{i} \,
\bigg[{1 \over 2} S (\Omega({\alpha}^{(i)})\,|\, {\rho}_{\Omega})
+ {1 \over 2}  S (\Omega({\beta}^{(i)}) \,|\, {\rho}_{\Omega}) \bigg]
\ee
Now since ${\rho}_{\Omega}$ is the optimal output state for the channel $\Omega$,
it is also the state which achieves the infimum in
the OPWSW representation (\ref{OPWSW}). Hence (\ref{OPWSW}) implies
\be
S (\Omega({\alpha}^{(i)})\,|\, {\rho}_{\Omega}) & \leq & {\chi}^{*}(\Omega) \\ \nonumber
S (\Omega({\beta}^{(i)})\,|\, {\rho}_{\Omega}) & \leq & {\chi}^{*}(\Omega)
\ee
Substituting into (\ref{suff5}) establishes (\ref{suff4}),
and hence the inequality (\ref{eqn4}), which proves the result.

\section{Proof of Theorem \ref{thm3}}
Theorem \ref{thm3} follows by combining Theorem \ref{thm2} with symmetry
and convexity arguments. We present the latter in this section, and postpone
the proof of Theorem \ref{thm2} to the next section.

\medskip
Our symmetry and convexity arguments use the classification of unital
qubit maps which was developed in \cite{KR}, and which we now review.
Any unital qubit map $\Phi$ can be represented by a real $3 \times 3$ matrix
with respect to the basis ${\sigma}_{1}, {\sigma}_{2}, {\sigma}_{3}$,
where ${\sigma}_{i}$ are the Pauli matrices. In \cite{KR} it was explained
that by using independent unitary transformations in
its domain and range, this matrix can be put into the 
following diagonal form:
\be\label{Phi}
\Phi = \left(\matrix{
{\lambda}_{1} & 0 & 0 \cr
0 & {\lambda}_{2} & 0 \cr
0 & 0 & {\lambda}_{3} \cr} \right)
\ee
The diagonal entries satisfy $|{\lambda}_{i}| \leq 1$, as well as other
conditions implied by complete positivity (these are described below).
The quantities ${\nu}_{p}$, $S_{\rm min}$ and ${\chi}^{*}$ are 
invariant under permutations of the
coordinates. They are also unchanged if the signs of any two of the parameters
${\lambda}_{1}, {\lambda}_{2}, {\lambda}_{3}$ are simultaneously flipped, as
this is implemented by a unitary transformation in the domain of
$\Phi$ (for example, conjugation by ${\sigma}_1$ in
the domain of $\Phi$ switches the signs of ${\lambda}_{2}$ and ${\lambda}_{3}$
without any other changes). So without loss of generality we will assume henceforth
that the parameters satisfy
\be\label{std}
1 \geq {\lambda}_{3} \geq \max ( |{\lambda}_{1}|, |{\lambda}_{2}|)
\ee
We will say that $\Phi$ is in {\it standard form} if it is diagonal in the basis
${\sigma}_{1}, {\sigma}_{2}, {\sigma}_{3}$ and its diagonal entries satisfy
(\ref{std}).

Assume that $\Phi$ is in standard form, and let
\be
\lambda = {\lambda}_{3} = \max ( |{\lambda}_{i}| )
\ee
Recall the definition (\ref{def:m_p}). Then
it is an easy calculation to show that
\be\label{eval}
{\nu}_{p}(\Phi) = m_{p}(\lambda)
\ee
For any unitary $2 \times 2$ matrix $U$ we define the qubit channel
${\Gamma}_{U}$ to be conjugation by the matrix $U$, so it acts on
a qubit state $r$ by
\be\label{def:Gamma}
{\Gamma}_{U} (r) = U \, r \, U^{*}
\ee
\medskip

\begin{lemma}\label{lemma1}
Let $\Phi$ be a unital qubit map in standard form, and let
$\lambda = {\lambda}_{3} = \max ( |{\lambda}_{i}| )$. 
Then $\Phi$ is a convex
combination of channels of the form
${\Gamma}_{W_{i}} \circ {\Psi}_{\lambda} \circ {\Gamma}_{U_{i}}$,
where ${\Psi}_{\lambda}$ is the phase-damping channel (\ref{ph-damp}) and 
$W_{i}, U_{i}$ are unitary matrices. Furthermore,
let $r$ be any qubit state. Then the unitary matrices $\{U_{i}\}$
can be chosen so that for each $i$,
\be\label{lemma:cond}
\tr \, {\sigma}_{3} ( U_{i} \, r \, U_{i}^{*} ) = 0
\ee
\end{lemma}

\bigskip
Before proving this lemma, we use it to deduce Theorem \ref{thm3}.
Let $\Phi$ be a unital qubit channel, and
let $\rho$ be a state on ${\bf C}^K \otimes {\bf C}^2$.
Let $r = {\tr}_{1}(\rho)$ be the $2 \times 2$ reduced density
matrix of $\rho$. Lemma \ref{lemma1} implies that there are constants
$c_{i} \geq 0$ with $\sum c_{i} = 1$, such that
\be
\Phi = \sum_{i} c_{i} \, {\Gamma}_{W_{i}} \circ 
{\Psi}_{\lambda} \circ {\Gamma}_{U_{i}}
\ee
and where the unitary matrices $\{U_{i}\}$ can be chosen so that
condition (\ref{lemma:cond}) is satisfied for this matrix $r$.
Using these same unitary matrices
$W_{i}, U_{i}$ and constants $c_i$ we can write
\be\label{convex.comb}
(I \otimes \Phi) (\rho) = 
\sum_{i} c_{i} \, (I \otimes W_{i}) \, (I \otimes {\Psi}_{\lambda})  
\big( {\rho}^{(i)} \big) \, (I \otimes W_{i}^{*})
\ee
where we have defined
\be\label{def:rho^i}
{\rho}^{(i)} = (I \otimes U_{i})  \, \rho  \,(I \otimes U_{i}^{*})
\ee
From (\ref{convex.comb}) we deduce
\be
||(I \otimes \Phi) (\rho) \big)||_{p} \leq
\sum_{i=1}^{N} c_{i} \,
|| \big(I \otimes {\Psi}_{\lambda}\big) \, \big( {\rho}^{(i)} \big) ||_{p}
\ee
Applying Theorem \ref{thm2} in the form (\ref{restate}) immediately gives
the statement of Theorem \ref{thm3}.

\bigskip
\bigskip
Now we present the proof of Lemma \ref{lemma1}. 
As was shown in \cite{KR} (and also in \cite{AF}), the allowed diagonal
entries of $\Phi$ in (\ref{Phi}) lie in the tetrahedron with corners
at the points 
\be\label{tetra}
(1,1,1), \,\, (1,-1,-1),\,\,
(-1,-1,1), \,\,(-1,1,-1)
\ee
For fixed $\lambda$,
the cross-section of this tetrahedron at height ${\lambda}_{3} = \lambda$
is a rectangle with corners at the four points
\be\label{4corners}
(1, \lambda, \lambda), \,\,(\lambda, 1, \lambda),\,\,
(-1, -\lambda, \lambda), \,\,(-\lambda, -1, \lambda)
\ee
It follows that if $\Phi$ is in standard form with 
$\lambda = {\lambda}_{3} = \max |{\lambda}_{i}|$, then $\Phi$ is a 
convex combination of the four maps corresponding to these corners.
Furthermore each of these maps is unitarily equivalent to the
phase-damping channel ${\Psi}_{\lambda}$ (\ref{ph-damp}). 
For example, the first map 
in (\ref{4corners}) acts on a state by
\be\label{corneracts}
r = \left(\matrix{x + y_{3} & y_{1} - i y_{2} \cr 
y_{1} + i y_{2} & x - y_{3}
}\right) \longmapsto
\left(\matrix{x + \lambda y_{3} & y_{1} - i \lambda y_{2} \cr 
y_{1} + i \lambda y_{2} & x - \lambda y_{3}
}\right),
\ee
and this same action can be written as
\be\label{def:V}
r \longmapsto V^{*} {\Psi}_{\lambda} (V \, r \, V^{*}) V
\ee
where $V$ is the unitary matrix $V = \exp [i ({\sigma}_{1} +
{\sigma}_{2} + {\sigma}_{3})/\sqrt{3} ]$. This unitary map
permutes the coordinates, that is
\be\label{permute}
V {\sigma}_{1} V^{*} = {\sigma}_{3}, \quad
V {\sigma}_{2} V^{*} = {\sigma}_{1}, \quad
V {\sigma}_{3} V^{*} = {\sigma}_{2},
\ee
and so the composition ${\Gamma}_{V^{*}} \circ {\Psi}_{\lambda} \circ
{\Gamma}_{V}$ reproduces the action of (\ref{corneracts}).
This establishes the first claim in Lemma \ref{lemma1}.

In order to derive the condition (\ref{lemma:cond}) we must
look more closely at the constraints on the unital maps.
As was shown in \cite{KR},
the condition (\ref{std}), namely
$\lambda = {\lambda}_{3} = \max ( |{\lambda}_{i}| )$,
selects a convex subset of the cross-section
of the rectangle (\ref{4corners}). 
For $1/3 \leq \lambda \leq 1$
this subset is the convex hull of the six points
\be\label{corners1}
(\lambda, \lambda, \lambda), \,\,(2 \lambda - 1, \lambda, \lambda),\,\,
({\lambda}, 2 \lambda - 1, \lambda), \cr
(-\lambda, -\lambda, \lambda), \,\,(1 - 2 \lambda, -\lambda, \lambda),\,\,
(-{\lambda}, 1 - 2 \lambda, \lambda),
\ee
and for $0 \leq \lambda \leq 1/3$ 
this subset is the convex hull of the four points
\be\label{corners2}
(\lambda, \lambda, \lambda), \,\,(-\lambda, \lambda, \lambda),\,\,
({\lambda}, -\lambda, \lambda),\,\,
(-\lambda, -\lambda, \lambda)
\ee
Hence it is enough to establish (\ref{lemma:cond}) for the
corner maps in (\ref{corners1}) and (\ref{corners2}). Furthermore,
the last three maps in (\ref{corners1}) can be transformed into
the first three by a unitary conjugation in the domain, and
the second and third maps are related by a permutation of coordinates,
so it is sufficient to consider just the first two maps in (\ref{corners1}).
Similarly it is sufficient to consider just the first two maps in
(\ref{corners2}).

\medskip
The key idea now is to use the additional symmetry of the corner maps 
(\ref{corners1}) and (\ref{corners2}) to arrange for the condition
(\ref{lemma:cond}) to be satisfied.
Consider the map corresponding to the first corner $(\lambda, \lambda, \lambda)$ in
(\ref{corners1}). This is the
well-known depolarizing channel ${\Delta}_{\lambda}$, and 
the symmetry of this map implies that for any state $r$ and any unitary
matrix $U$,
\be\label{pullthrough}
U \,{\Delta}_{\lambda} (r)\, U^{*} = {\Delta}_{\lambda} (U\, r\, U^{*})
\ee
The idea now is to choose $U$ to diagonalize $r$. That is, choose
$U$ so that
\be
U \,r \,U^{*} = \left(\matrix{{1 \over 2} + y & 0 \cr 
0 & {1 \over 2} - y
}\right)
\ee
where $|y| \leq 1/2$. This means in particular that
\be\label{cond2}
\tr \,{\sigma}_{1} (U\, r\, U^{*}) = \tr \,{\sigma}_{2} (U\, r\, U^{*}) = 0
\ee
It follows that we can write
\be\label{change}
{\Delta}_{\lambda} (r) = U^{*} \,
{\Delta}_{\lambda} (U\, r \,U^{*})\, U
\ee
Now we write ${\Delta}_{\lambda}$ as a convex combination of the four corners
(\ref{4corners}) (this can be done in many ways, the precise choice does
not matter). Having done this, it is enough to establish Lemma \ref{lemma1}
for each of these four corner maps applied to the state $U\, r \,U^{*}$. For example, 
the first map in (\ref{4corners}) acts according to (\ref{def:V}), and so
when applied to the state $U\, r \,U^{*}$ it gives
\be
 V^{*} \,{\Psi}_{\lambda} (V\, U \,r\, U^{*} \,V^{*} ) \,V
\ee
The action of $V$ in (\ref{permute}) together with (\ref{cond2}) imply
that
\be
\tr \,{\sigma}_{3} (V\, U\, r \,U^{*} \,V^{*} ) = 
\tr \, {\sigma}_{1} (U\, r \,U^{*} ) =
0
\ee
Hence the condition (\ref{lemma:cond}) is satisfied for this term.
The other three corners in (\ref{tetra}) produce similar
expressions, and so we have written ${\Delta}_{\lambda} (r)$ as a convex
combination of terms of the form
$W_{i} \, {\Psi}_{\lambda} (U_{i} \, r \, U_{i}^{*}) \, W_{i}^{*}$
with $\tr \, {\sigma}_{3} \, (U_{i} \, r \, U_{i}^{*}) = 0$, which
establishes (\ref{lemma:cond}) for the depolarizing channel.

\medskip
The map corresponding to the second corner in
(\ref{corners1}), namely $(2 \lambda - 1, \lambda, \lambda)$, is the
two-Pauli channel of Bennett, Fuchs and Smolin \cite{BFS}.
The analysis for this channel follows the same lines as for the depolarizing channel,
so we just explain the steps here. 
First, given a state $r$, use the $y-z$ symmetry of the channel to `pull through' a unitary
transformation $\exp [i \theta {\sigma}_{1}]$ onto $r$
(in the same way as for the depolarizing channel in (\ref{pullthrough})), 
and choose $\theta$
so that the resulting state $r'$ satisfies
\be\label{trace0}
\tr \,{\sigma}_{2} r' = 0
\ee
Next, write the channel $(2 \lambda - 1, \lambda, \lambda)$ as a
convex combination of $(\lambda, 1, \lambda)$ and
$(-\lambda, 1 - 2 \lambda, \lambda)$ (recall that 
$1/3 \leq \lambda \leq 1$ so that the latter is within the tetrahedron
of unital maps (\ref{tetra})).
The first map $(\lambda, 1, \lambda)$ 
is one of the corners (\ref{4corners}), and it is applied
to the state $r'$ which satisfies (\ref{trace0}), hence by a unitary
transformation which permutes ${\sigma}_{1} \rightarrow
{\sigma}_{2} \rightarrow {\sigma}_{3}$ we can rewrite it as the action
of the phase-damping channel on a state satisfying 
(\ref{lemma:cond}). The second channel $(-\lambda, 1 - 2 \lambda, \lambda)$
is unitarily equivalent to $(\lambda, 2 \lambda - 1, \lambda)$, which is
again a two-Pauli channel. Using the symmetry of this channel, we can pull
through a unitary transformation $\exp [i \phi {\sigma}_{2}]$ onto the state
$r'$. This does not affect the condition (\ref{trace0}), and so
by choosing $\phi$ correctly the resulting state $r''$ satisfies
\be\label{trace1}
\tr {\sigma}_{1} r'' = \tr {\sigma}_{2} r'' = 0
\ee
Now we can write the channel $(\lambda, 2 \lambda - 1, \lambda)$ as a 
convex combination of the four corners (\ref{4corners}), and repeat the
argument for the depolarizing channel, concluding that each of these corner 
maps acting on $r''$ is unitarily
equivalent to the phase-damping channel acting on a state
satisfying (\ref{lemma:cond}). This establishes the result for the two-Pauli
channel.

\medskip
Finally consider the case $1/3 \geq \lambda \geq 0$, where the corner maps are
(\ref{corners2}). The first map is the depolarizing channel ${\Delta}_{\lambda}$,
which was done above. The second map is unitarily equivalent to the
depolarizing channel ${\Delta}_{-\lambda}$, and so this is also done.

\section{Proof of Theorem \ref{thm2}}
This theorem is a variant of the bound obtained by Lieb and Ruskai, which appeared
as an Appendix in the paper \cite{K}, and it can be proved by the same method.
That method uses the Lieb-Thirring bound \cite{LT}, which in turn was proved using
one of Epstein's concavity results \cite{Ep}. Since Theorem \ref{thm2} can be obtained
directly from Epstein's result, we present that argument here.

Let $\rho$ be a matrix of the form (\ref{rho}). The condition that $\rho$
be positive means that $Y_{1} - i Y_{2} = \sqrt{X+Y_{3}}\, R \,\sqrt{X-Y_{3}}$ where
$R$ is a contraction. Every contraction is a convex combination of
unitaries, so it is sufficient to assume that
\be
Y_{1} - i Y_{2} = \sqrt{X+Y_{3}}\, V \,\sqrt{X-Y_{3}}, \quad\quad V V^{*} = I
\ee
We have the factorization
\be
(I \otimes {\Psi}_{\lambda}) (\rho) = F^{1/2} \, G \, F^{1/2}
\ee
where
\be
F = \left(\matrix{{X+Y_{3}} & 0 \cr 0 & {X-Y_{3}} }\right), \quad\quad
G = \left(\matrix{I & \lambda V \cr \lambda V^{*} & I }\right)
\ee
From the identity
\be
\left(\matrix{I & \lambda V \cr \lambda V^{*} & I }\right)  = 
U\,\left(\matrix{(1+\lambda) I & 0 \cr 0 & (1-\lambda) I }\right) \,U^{*},
\ee
where
\be
U  =  {1 \over \sqrt{2}}
  \pmatrix{ I &  V \cr  V^* & -I} ,\quad\quad U U^{*} = I,
\ee
it follows that for all $p \geq 1$
\be\label{G^p}
G^{p} = \left(\matrix{\alpha I & \beta V \cr \beta V^{*} & \alpha I }\right)
\ee
with
\be
\alpha = {1 \over 2} [(1+\lambda)^p + (1-\lambda)^p ],\quad\quad
\beta = {1 \over 2} [(1+\lambda)^p - (1-\lambda)^p ]
\ee
We can write
\be\label{eq1}
\tr \bigg( (I \otimes {\Psi}_{\lambda}) (\rho) \bigg)^p = 
\tr \bigg( F^{1/2} \, ( G^p )^{1/p} \, F^{1/2} \bigg)^p
\ee
Now we use Epstein's concavity result \cite{Ep}, which states that for any positive matrix
$B$ and any $p \geq 1$, the map
\be\label{Ep}
A \,\, \rightarrow \tr \bigg( B \, (A)^{1/p} \, B \bigg)^p
\ee
is concave on the set of positive matrices. (In fact Epstein states the result
only for integer values of $p$, but his proof applies to all real values
$p \geq 1$). The left side of (\ref{eq1}) is an even function of $\lambda$,
and therefore the right side is unchanged if $\beta$ is replaced by $-\beta$
in (\ref{G^p}). Also note that
\be
{1 \over 2} \, \left(\matrix{\alpha I & \beta V \cr \beta V^{*} & \alpha I }\right)
+ {1 \over 2} \, \left(\matrix{\alpha I & -\beta V \cr -\beta V^{*} & \alpha I }\right)
= \left(\matrix{\alpha I & 0 \cr 0 & \alpha I }\right)
\ee
Therefore the concavity result (\ref{Ep}) implies that the right side
of (\ref{eq1}) is bounded above by its value when $\beta$ is set equal to
zero  in (\ref{G^p}). Furthermore when $\beta = 0$, the right side of (\ref{eq1}) becomes
\be
\tr \, \bigg( F^{1/2} \, \big( \alpha \,I\big)^{1/p} \, F^{1/2} \bigg)^p = 
\alpha \, \tr \, F^p = 
2 \, \alpha \, 
\bigg[ {1 \over 2} \tr \big( X + Y_{3} \big)^{p}
+ {1 \over 2} \tr \big( X - Y_{3} \big)^{p}
\bigg]
\ee
Comparing with (\ref{def:m_p}) we see that
$2 \, \alpha = 2^{p} \, {m_{p}(\lambda)}^{p}$, and this
proves the theorem.

\bigskip
{\bf Acknowledgements}
The author is grateful to M. B. Ruskai for helpful discussions and
comments.

{~~}

\end{document}